# A capacitive flexible tactile sensor

Dandan Yuan, Haoxin Shu*, Yulong Bao, Bin Xu*, Huan Wang

Jiangxi Province Key Laboratory of Precision Drive & Control, Nanchang Institute of Technology, Nanchang 330099, China

xubin84115@163.com

## Abstract

In this paper, a capacitive flexible tactile sensor was designed to measure the pressure of objects based on MEMS technology. This sensor is a structure of a 4x4 array, with metal Ag as the capacitive electrode, which forms the tactile sensing unit of the sensor. The structure of capacitive flexible tactile sensor was designed and an experimental platform was established to test the performance. The tests show that when the thickness of the intermediate layer is 2 mm and the density is medium, the sensor's sensitivity is the best while the time of both the response and the rebound is fast.

**Keywords:** tactile sensors, flexible, MEMS, capacitive, pressure testing

## 1. Introduction

With the rapid development of society, a research hotspot field of robotics acttracts more and more attention. The flexible tactile sensor is self-evidently more and more important to a robot in terms of sensing the pressure of external objects Many scholars at home and abroad have conducted researches and achieved some results. HU XiaoHui et al [1] developed a flexible capacitive tactile array sensor with the microneedle structure in the robotics field. This device has better flexibility because PDMS microneedle layer is sandwiched between the upper and lower electrodes. M.-Y. Cheng et al [2] designed a capacitive touch array sensor which is composed of a PDMS structure and a flexible printing plate with electrodes. It effectively reduces the complexity of the capacitor structure of each sensing element. In 2019, Jie Qiu and others prepared a flexible capacitive tactile sensor with fast response, low detection limit, and high sensitivity [3]. However, because the response and rebound time of a flexible tactile sensor is also an important performance metric, a capacitive flexible tactile sensor with this structure is proposed in this paper. According to the working principle, tactile sensors can be divided into piezoresistive type [4], piezoelectric type [5], capacitive type [6-7], optical sensor [8], magnetic sensor [9], etc. Capacitive tactile sensors have received wide attention and have been extensively used benefiting from their good linear response, wide dynamic range, and favorable durability [10]. This study has made three kinds of intermediate medium layers with different thickness and density. The best one is obtained by the sensitivity test.

## 2. Structure Design

The overall three-dimensional structure of the new capacitive flexible tactile sensor is shown in Figure 1. The sensor is a structure of a 5-layer device, including a contact layer, an upper electrode plate and an upper electrode, an intermediate dielectric layer, a lower electrode plate and a lower electrode.

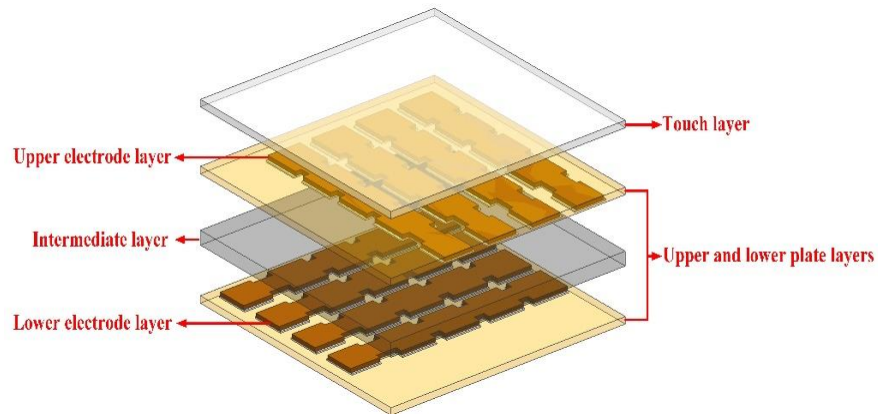

Figure 1. Structure diagram of capacitive flexible tactile sensors

The designed capacitive flexible tactile sensor uses PDMS film as the contact layer. PDMS is a kind of polymer organic silicide. It is able to withstand multiple touch and press, effectively protects the capacitive flexible tactile sensor from damage, and improves the durability of multiple touch and press. The upper and lower plates are made of PI material. It is a flexible polymer material that has good heat resistance, it has good corrosion resistance, excellent electrical insulation, good chemical stability and so on. The upper and lower electrodes are made of metal Ag, an excellent sensor electrode material which has low activity, good thermal conductivity and electrical conductivity. Ag is soft, ductile and not easily corroded by chemicals. The intermediate medium layer is made of PU material (polyurethane filter sponge) whose main body is a mesh structure. So it has excellent air permeability, good flexibility, and is durable, environmentally friendly and low-priced.

The overall size of the upper and lower electrode layers is a square of 24mm×24mm, and the electrodes are made into a 4×4 sensor array. The upper and lower electrodes are staggered horizontally and vertically, and each intersecting place constitutes a sensing unit, so there are 16 independent sensing units in total. The structure of the electrode layer is shown in Figure 2.

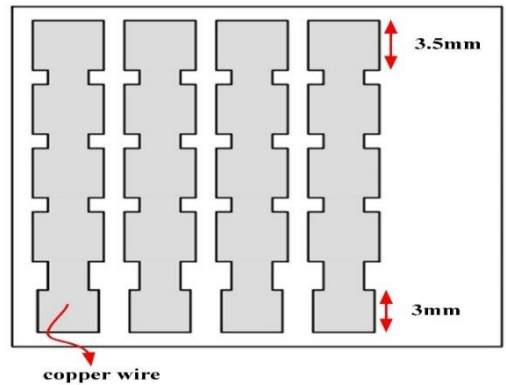

Figure 2. Electrode layer structure diagram

## 3. Preparation Process

The fabrication process of a capacitive flexible tactile sensor is shown in Figure 3. It shows the process flow of the four layers using PI flexible substrate MEMS technology.

(a)Silicon wafer is used as the bottom plate with a PI tape pasted on it. Then, it is cleaned in sequence with acetone, anhydrous ethanol and deionized water. After the cleaning, it is put on the heating platform for drying.

(b) A layer of positive photoresist is uniformly sprayed on the surface of PI film. The negative film is fixed on the disc by vacuum pump. When the negative film is rotating at a high speed, the photoresist is evenly distributed on the negative film by the glue throwing machine.

(c) The photoetching process is carried out by a photoetching machine. The photoetching machine exposes the photoresist through an ultraviolet light source and puts the exposed film into the developer. The photoresist in the exposed area is etched away and the photoetching pattern is fully displayed. Then it is rinsed in deionized water and then taken out for natural air drying.

(d) The sample is sputtered with a high vacuum triple target magnetron coater. The vacuum is pumped to $10^{-4}$Pa, the flow rate controlled to 80, and the power controlled to 100 W. After sputtered with Cr target for three minutes and then with Ag target for 30 minutes, the sample is cooled and taken out.

(e) The purpose of stripping process is to remove the remaining photoresist, so that the electrode pattern can be displayed. After the remaining photoresist and Ag attached to the photoresist are removed with acetone, the Ag electrode remained.

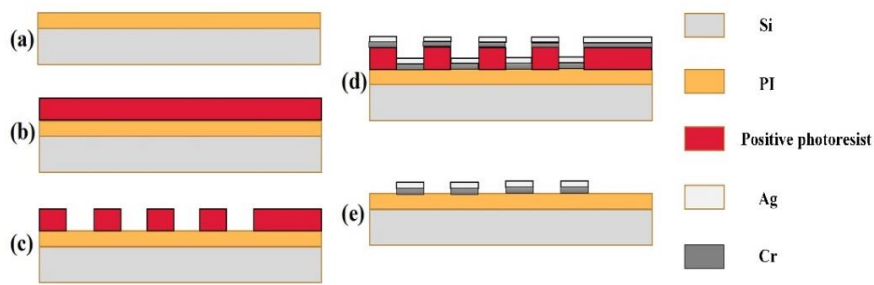

Figure 3. Electrode layer process flow

## 4. Experiment and Analysis

The experimental platform of the capacitive tactile sensor is shown in *Figure 4(a)*. It consists of the LCR digital bridge tester and digital display push-pull tester. The capacitive flexible tactile sensor is fixed on a circular plate so that its circular contact is just above the contact element. A certain pressure is applied through moving the push-pull gauge vertically downward by hand cranking. The physical picture of the device is shown in Figure 4(b).

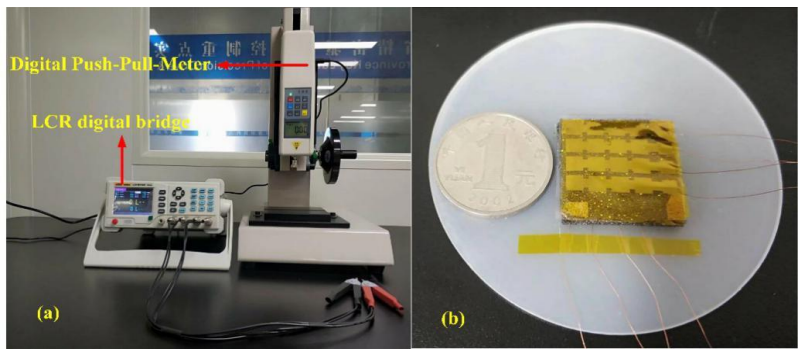

Figure 4. (a) Experimental Platform (b) Fabricated flexible tactile sensors

In this paper, we have made three kinds of PU materials for the intermediate dielectric layer. They are ones of 2mm thick and medium density, 3mm thick and medium density, and 3mm thick and high density respectively. Because a single different parameter need to be set for comparison, we compare the experimental results of different intermediate dielectric layers with medium density and high density with the thickness of 3mm PU. Figure 5(a) shows the test results of two capacitive flexible tactile sensors. The results show that for the same thickness of 3mm, the initial value of high density is 0.53pf, the sensitivity 12.58% / N, the initial value of medium density 1.45pf, and the sensitivity 15.98% / n. In comparison with the experimental results of the same medium density, 2mm thick and 3mm thick, as shown in Figure 5(b), for the same medium density, the initial value of 2mm thickness is 1.4pF and the sensitivity is 22.62%/N. Therefore, through the comparison of three different types of capacitive tactile sensors, the sensitivity of the sensor is the highest when the thickness of the intermediate dielectric layer is 2mm and the density is medium density.

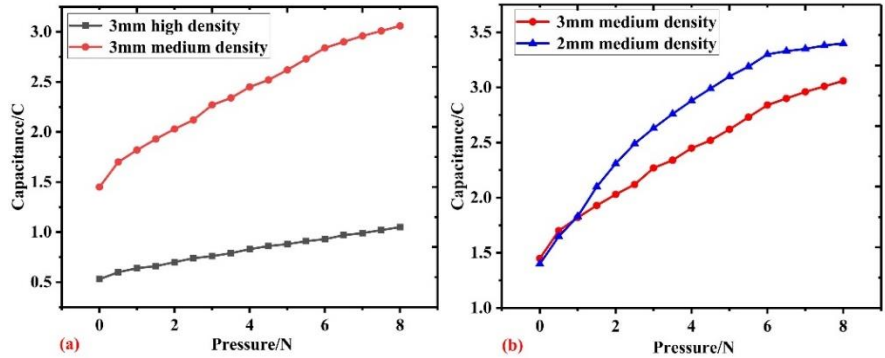

Figure 5. (a) 3mm medium and high-density capacitance value comparison (b) 2mm and 3mm medium-density thickness capacitance comparison

When the applied pressure reaches 6N and the intermediate dielectric layer is 2mm thick, which indicates that the capacitive sensor of this model has reached the maximum range. The sensitivity of the capacitive sensor can be expressed as:

$$K = \frac{\Delta C/C_0}{\Delta N} \tag{1}$$

Through date analysis, when the applied pressure is 6N, the sensor with 3mm high density has the lowest sensitivity of 12.58% / N, and the sensor with 2mm medium density has the highest sensitivity of 22.62% / N.

The response time and rebound time of the capacitive touch sensor to the pressure signal are also very important in application, which can reflect the performance of the capacitive touch sensor. In this paper, response time and rebound time of a 2mm medium-density capacitive tactile sensor are tested. The pressure of 1N, 2N, 3N, 4N is applied in sequence to a tactile unit for 2s, and the time for releasing the pressure at intervals is 3s and a set of data is taken every 0.1s. The experimental results are shown in Figure 6. It can be seen from the figure that the response time of the capacitive touch sensor is 0.1s, and the rebound time is 0.3s. Therefore, the response time of the capacitive touch sensor is faster.

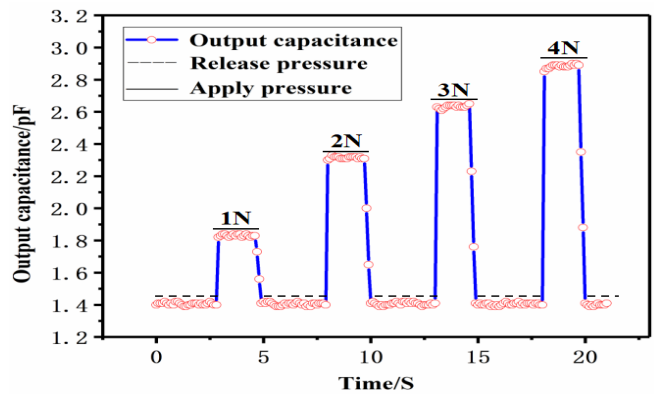

Figure 6. Capacitive tactile sensor response time and rebound time

## 5. Conclusion

In this paper, a capacitive flexible tactile sensor is fabricated by MEMS technology. Three sensors with different intermediate dielectric layers has been designed for comparison. It studies the relationship between the sensitivity and the thickness and density parameters of capacitive flexible tactile sensors, as well as the range, response time and rebound time. After many experimental tests, it can be seen that the comprehensive performance of the intermediate medium layer with thickness of 2mm and medium density is the best.

## Acknowledgments

This work was supported in part by Science Foundation of the department of education of Jiangxi province (GJJ180938, GJJ170987, GJJ180936). Open project of Shanxi Key Laboratory of Intelligent Robot(SKLIRKF2017004).

# References

[1] HU XiaoHui,ZHANG Xu,LIU Ming,CHEN YuanFang,LI Peng,PEI WeiHua,ZHANG Chun,CHEN HongDa.A flexible capacitive tactile sensor array with micro structure for robotic application[J].Science China
(Information Sciences),2014,57(12):41-46.

[2] Cheng M Y, Liao B T, Huang X H, et al. A flexible tactile sensing array based on novel capacitance mechanism[C]//TRANSDUCERS 2009-2009 International Solid-State Sensors, Actuators and Microsystems Conference. IEEE, 2009: 2182-2185.

[3] Qiu J, Guo X, Chu R, et al. Rapid-Response, Low Detection Limit, and High-Sensitivity Capacitive Flexible Tactile Sensor Based on Three-Dimensional Porous Dielectric Layer for Wearable Electronic Skin[J]. ACS Applied Materials & Interfaces, 2019, 11(43): 40716-40725.

[4] Fu Y M, Chou M C, Cheng Y T, et al. An inkjet printed piezoresistive back-to-back graphene tactile sensor for endosurgical palpation applications[C]//2017 IEEE 30th International Conference on Micro Electro Mechanical Systems (MEMS). IEEE, 2017: 612-615.

[5] Liu W, Yu P, Gu C, et al. Fingertip piezoelectric tactile sensor array for roughness encoding under varying scanning velocity[J]. IEEE Sensors Journal, 2017, 17(21): 6867-6879.

[6] Hu X H, Zhang X, Liu M, et al. A flexible capacitive tactile sensor array with micro structure for robotic application[J]. Science China Information Sciences, 2014, 57(12): 1-6.

[7] Sun C T, Lin Y C, Hsieh C J, et al. A linear-response CMOS-MEMS capacitive tactile sensor[C]//SENSORS, 2012 IEEE. IEEE, 2012: 1-4.

[8] Jiang Q, Xiang L. Design and experimental research on small-structures of tactile sensor array unit based on fiber Bragg grating[J]. IEEE Sensors Journal, 2017, 17(7): 2048-2054.

[9]Volkova T I, Böhm V, Naletova V A, et al. A ferrofluid based artificial tactile sensor with magnetic field control[J]. Journal of Magnetism and Magnetic Materials, 2017, 431: 277-280.

[10]Zou L, Ge C, Wang Z J, et al. Novel tactile sensor technology and smart tactile sensing systems: A review[J]. Sensors, 2017, 17(11): 2653.